\documentclass[aps,prb,twocolumn,floatfix,showkeys,superscriptaddress,amssymb,preprintnumbers]{revtex4-1}
\usepackage{graphicx}
\usepackage{bm}                   
\usepackage{amsmath}
\usepackage{csquotes}        
\usepackage[linktocpage, colorlinks, linkcolor = blue, citecolor = blue]{hyperref}
\usepackage{dcolumn}        
\usepackage[caption=false]{subfig}
\usepackage[dvipsnames]{xcolor}				

\newcommand{\vv}{{\rm v}}
\newcommand{\gb}{{\rm gb}}
\newcommand{\dd}{{\rm d}}

\newcommand{\m}{{\rm m}}

\newcommand{\an}{{\rm an}}
\newcommand{\eff}{{\rm eff}}

\newcommand{\mm}{{\rm m}}
\newcommand{\pp}{{\rm p}}
\newcommand{\av}{{\rm av}}
\newcommand{\Co}{{\rm Co}}
\begin{document}

\title{Grain boundary diffusion in severely deformed Al-based alloy}


\author{Sergiy V. Divinski}\email{For correspondence: divin@wwu.de}
\affiliation{Institute of Materials Physics, University of M\"unster, D-48149 M\"unster, Germany}
\author{Vladislav Kulitcki}
\affiliation{Institute of Materials Physics, University of M\"unster, D-48149 M\"unster, Germany}
\affiliation{Belgorod State University, Belgorod, 308015, Russia}
\author{Beng\"u Tas Kavakbasi}
\affiliation{Institute of Materials Physics, University of M\"unster, D-48149 M\"unster, Germany}
\author{Ankit Gupta}
\affiliation{Max-Planck-Institut f\"ur Eisenforschung GmbH, D-40237 D\"usseldorf, Germany}
\author{Yulia Buranova}
\affiliation{Institute of Materials Physics, University of M\"unster, D-48149 M\"unster, Germany}
\author{Tilmann Hickel}
\affiliation{Max-Planck-Institut f\"ur Eisenforschung GmbH, D-40237 D\"usseldorf, Germany}
\author{J\"org Neugebauer}
\affiliation{Max-Planck-Institut f\"ur Eisenforschung GmbH, D-40237 D\"usseldorf, Germany}
\author{Gerhard Wilde}
\affiliation{Institute of Materials Physics, University of M\"unster, D-48149 M\"unster, Germany}
\date{\today}

\begin{abstract}
	
Grain boundary diffusion in severely deformed Al-based AA5024 alloy is investigated. Different states are prepared by combination of equal channel angular processing and heat treatments, with the radioisotope $^{57}$Co being employed as a sensitive probe of a given grain boundary state. Its diffusion rates near room temperature (320~K) are utilized to quantify the effects of severe plastic deformation and a presumed formation of a previously reported deformation-modified state of grain boundaries, solute segregation at the interfaces, increased dislocation content after deformation and of the precipitation behavior on the transport phenomena along grain boundaries. The dominant effect of nano-sized Al$_3$Sc-based precipitates is evaluated using density functional theory and the Eshelby model for the determination of elastic stresses around the precipitates.
\end{abstract}

\keywords{Al 5024 alloy; Severe plastic deformation; Grain boundary diffusion; Precipitation}
\maketitle

\section{Introduction}

\noindent Al-based alloys are attractive materials for various industrial applications owing to their low density combined with sufficient mechanical strength and ductility. The severe plastic deformation (SPD) of metals is a method to improve the mechanical properties of the material via a strong microstructure refinement and the formation of nanostructures \cite{V1, V2}. Equal channel angular pressing (ECAP) is one such deformation technique beside of high pressure torsion (HPT) or accumulative roll bonding (ARB) that allows producing materials with ultra-fine grain (UFG) microstructure and supports industrial up-scaling \cite{rev}. ECAP processing at elevated temperatures was, e.g., applied to achieve superplastic properties of Al-based alloys \cite{Mog}.

The material modification via SPD processing is not only limited to grain refinement, but also includes modifications of the grain boundary (GB) kinetics, structure and segregation \cite{Sauvage} simultaneously with the production of abundant point defects and dislocations in the processed material \cite{Zehet, EV}. For example, an enhancement of grain boundary self-diffusion rates by orders of magnitude was established for ECAP-processed Ni \cite{Gerrit} or Ti \cite{Jochen}, a fact which was unambiguously attributed to a deformation-modified (in pioneering works termed as \enquote{non-equilibrium} \cite{Nazarov}) state of the interfaces due to defect incorporation and enhancement of the excess free volume \cite{IF, Jochen}. In some cases, an extra-ordinary high strength of UFG materials was attributed to specific grain boundary segregations \cite{Nariman}. 

It is well known that beyond a critical strain, a further increase of the applied total strain does not produce further grain refinement, and a saturation of the microstructure modification is observed which corresponds to a balance between the rates of defect generation and their annihilation \cite{EV}. For the given deformation temperature and strain rate, the relative rate of recovery processes will be facilitated in the case of materials with a lower melting point due to a known semi-empirical relationship between the diffusion rates and the melting temperatures of various classes of materials \cite{Mehrer}.

As stated above, the deformation-modified state of grain boundaries was documented for Ni and Ti with relatively high melting points deformed at room temperature. Does the SPD processing of low-melting point materials like Al or Al-based alloys induce the deformation-modified grain boundaries with enhanced diffusion rates, too? Cold plastic deformation of Al has been shown \cite{Grabski, Grabski1} to modify the interface structures in pure Al by introducing so-called extrinsic grain boundary dislocations. Still these modifications were limited exclusively to special grain boundaries in Al, whereas the general high-angle GBs relaxed too fast already at room temperature for any modification to be observed. Addition of Mg to Al stabilized extrinsic grain boundary dislocations and a deformation-modified state of interfaces was stated to have been observed which then relaxed at elevated temperatures \cite{Gleiter, Horita}. For example, zigzag, highly stepped configurations of grain boundaries in an UFG Al--3\% Mg alloy which recovered after irradiation by the electron beam in TEM were reported \cite{Horita}. Although these observations were discussed in terms of the deformation-modified ('non-equilibrium' as it was originally introduced \cite{Gleiter, Grabski}) state of the interfaces \cite{Horita}, it is known that even well-annealed grain boundaries may also be faceted and dissociated on the atomic scale \cite{Baluffi}. Thus, kinetic measurements would be helpful to elucidate a hypothetical deformation-modified character of the grain boundaries in addition to the microstructure observations.

The present work is focused on studying the effect of severe plastic deformation on the kinetic properties of interfaces in an Al-based alloy and contributes to a better understanding of the coupling between the microstructure evolution, precipitate formation and local strain-stress fields as a function of annealing conditions. For these purposes, the tracer diffusion method is applied to follow the rate of atomic transport, which is extremely sensitive to the state of the grain boundaries, grain boundary segregation and grain boundary precipitation \cite{Divinski, Dandan}. The $^{57}$Co isotope is chosen as a convenient tracer element and its grain boundary diffusion is measured at about room temperature in Al-alloys which were subjected to different heat treatments at elevated temperatures. Thus, the determined diffusion coefficients represent a specific probe of the given grain boundary state on a time scale when bulk diffusion of all substitutional elements in Al is frozen and the attained state is not modified during low-temperature diffusion measurements.

A complex Al-based AA5024 alloy was chosen for the investigation, with Mg, Sc, and Zr as main alloying components. Sc was added on purpose to facilitate the grain refinement via nano-precipitation \cite{Sc, Yulia}. On the other hand, a strong segregation of Mg in this alloy is to be expected, too \cite{seg1, seg2}. As a result, a sophisticated interplay of hypothetical deformation-induced grain boundary modifications, grain boundary segregation and precipitation may be expected.

\section{Experimental procedure}

\subsection{Material}

\noindent The Al-based AA5024 alloy with the nominal composition  Al--4.6Mg--0.35Mn--0.2Sc--0.09Zr--0.2Ti--0.08Fe--0.02Si (in wt.\%) was prepared in the research group of Prof. R. Kaibyshev (Belgorod State University, Russia). It was produced by continuous casting and the ingots were homogenized by annealing at $(643 \pm 10)$~K for 12~h. The ingots were then extruded at $653$~K imposing about 70\% strain. The resulting rods were subjected to ECAP processing with application of about 100~MPa back-pressure at a temperature of 573~K \cite{Mog2}. The final grain size was about 500~nm. 

For a reference, coarse grained high-purity Al (99.999 wt.\%) was used. The material was annealed at 873~K for 16~h and slowly cooled. The grain size was about 0.5~mm.

\subsection{Sample preparation and characterization} 

\noindent Discs of 10~mm in diameter and 1.5~mm in height were cut by spark erosion. Some samples were subjected to pre-diffusion heat treatments at selected temperatures for the given times listed in Table~\ref{tab:par}. 

The microstructures were characterized by scanning electron microscopy (SEM) using a Nova Nano SEM 230 (FEI). Orientation imaging microscopy was applied using electron back-scatter diffraction (EBSD). For the EBSD analysis, the samples were mechanically ground, polished using diamond suspensions and finished with colloidal silica ($0.04$~$\mu$m) until a mirror-like surface finish was achieved.

Local microstructure analysis was performed with a transmission electron microscope (TEM) using a Libra 200 FE TEM (Zeiss) and a JEOL JEM-2100 (both with an acceleration voltage of 200~kV). The disk-shaped samples for TEM analysis were cut from sections with the foil-normal perpendicular to the extrusion direction in the central part of the processed materials. The samples from the ECAP-processed rods were cut perpendicularly to the extrusion direction. $1$~mm-thick disks, $3$~mm in diameter, were cut by spark erosion and mechanically polished down to the thickness of about $90$~$\mu$m. The final thinning was done by chemical electropolishing (twin-jet electro-polishing in a solution of HNO$_3$:CH$_3$OH with $1:2$ ratio at $253$~K). 

The sizes of the equiaxed grains were quantified by measuring the grain area and calculating an equivalent grain diameter by modeling each grain as a circle in accordance with the grain reconstruction method \cite{hump}. The size of the precipitates was estimated in bright-field TEM (BF-TEM) using at least ten arbitrarily selected micrographs, as it was employed in Ref.~\onlinecite{ma15}. Additionally, dark field-TEM (DF-TEM) and high-angle annular dark-field scanning TEM (HAADF-STEM) micrographs were employed to measure the size of about 100 precipitates in each state. The total number of individual measurements for each condition was about 1000, see also Refs. \onlinecite{m13, k05}. The dislocation density was calculated in TEM by measuring the number of dislocations for the given intersection length and foil thickness.

\begin{table*}
\caption{The results of the diffusion experiments for Co GB diffusion in pure Al and the Al-based AA5024 alloy at $320$~K for $3$ days, $D_{\gb}$. Before the diffusion experiments some samples were annealed at $T_{\an}$ for the time $t_{\an}$. The mean grain size, $d$, dislocation density, $\rho_\dd$, and the average particle size, $R_{\av}$, are specified. The parameter $\alpha^*$ is calculated using Eq.~(\ref{eq:a*}). The value of the GB diffusion coefficient measured for the ECAP-processed state (with $\alpha^* < 1$) has to be corrected and the corrected value is given, $D_{\gb}^{corr}$ (see also text).}\label{tab:par}
\begin{tabular}{c|cccccccc}																						\hline
 Sample & $T_{\an}$ (K) & $t_{\an}$ (h) & \parbox[c][0.8cm]{3cm}{$D_{\gb}$ ($10^{-15}$~m$^{2}$/s)} & $d$ ($\mu$m) & \parbox[c][0.8cm]{2.5cm}{$\rho_\dd$ ($10^{13}$~m$^{-2}$)} & $R_{\av}$ (nm) & \parbox[c][0.8cm]{1.5cm}{$\alpha^*$} & \parbox[c][0.8cm]{3cm}{$D_{\gb}^{corr}$ ($10^{-15}$~m$^{2}$/s)} \\
\hline
\vspace{3mm}
Pure Al & -- & -- & 17.5 & ~500 & \parbox[c][0.5cm]{3cm}{$10^{-2} - 10^{-1}$} & -- & 6000 & --\\
\vspace{3mm}
as-cast & -- & -- & 7.15 & 40   & $0.8$       & $5 - 6$ & 7.5  & --   \\
        & -- & -- & 7.21 & 0.8  & 7           & 11    & 0.85 & 9.61 \\
ECAP-   & 723& 1  & 4.16 & 1.2  & $<1$        & $12 - 14$ & $>6$ & -- \\
processed &773&20 & 1.41 & 7    & $<1$        & 20    & $>6$   & --\\
at 573 K& 823& 50 & 3.88 & 10   & $<1$        & $30 - 40$ & $>6$ & -- \\
        & 823&200 & 9.0  & $>20$& $<1$        & $>50$   & $>6$   & --\\
\hline
\end{tabular}

\end{table*}

The deformed microstructures were described in detail in previous publications \cite{Mog, Kul, Yulia}. Here we will present the most relevant details and the results for the microstructures relevant for subsequent diffusion measurements.

\subsection{Grain boundary tracer diffusion of $^{57}$Co}

\noindent Before the diffusion measurements, the surface of the samples was polished to a mirror-like quality. Three to five microliters of the $^{57}$Co tracer solution (272 days half-life and 122 keV $\gamma$-radiation), was dropped on the prepared surface and dried. The samples were sealed in silica tubes under a purified argon atmosphere and annealed. The annealing temperature was set at 320~K to be slightly above room temperature and the annealing time was fixed at 3 days. In order to eliminate the effects of surface and lateral diffusion, the sample diameter was reduced by about 1 mm by grinding after the diffusion annealing treatment.  

The parallel serial sectioning was performed using Mylar foils ($15~\mu$m particle size) on a custom-built precision grinding machine. After removing a section, the sample was weighed by a microbalance. The thickness of each section was determined by the mass difference from the known density and radius of the sample. The relative specific radioactivity of each section (which is proportional to the tracer concentration) was determined by measuring the radioactive decays of the $^{57}$Co tracer by a well-type intrinsic Ge $\gamma$-detector. 

The grain boundary diffusion conditions were set to fulfill Harrison's type C kinetic regime \cite{Har}. Therefore, the concentration profiles were plotted as the logarithm of the layer tracer concentration, $\ln \bar{c}$, vs. the depth squared, $y^2$, and the corresponding diffusion coefficients, $D_\gb$, were determined using the standard Gaussian solution \cite{PD},

\begin{equation}
 D_\gb = \frac{1}{4t} \left( - \frac{\partial \ln \bar{c}}{\partial y^2} \right)^{-1} \label{eq:Dgb}
\end{equation}

\noindent Here $t$ is the diffusion time.

\section{Density  Functional Theory (DFT) METHODOLOGY}

\noindent The present paper is focused on the effect of elastic strain in the Al-based matrix on GB diffusion of Co. However, direct DFT computations for general high-angle GBs (as they are addressed in the experimental part) represent an extremely involved task and a simplified approach is used here. We will make use of the experimentally established correlation between the activation energies of GB diffusion, $Q_\gb$, (measured directly in the corresponding C-type regime for solute GB diffusion) and those for bulk diffusion in the same matrix, $Q_\vv$, in metals, $Q_\gb \approx (0.4-0.6) Q_\vv$ \cite{Mehrer, PD}. Moreover, the pressure-dependent measurements of Zn GB diffusion in Al provide similar values of the activation volume which could be expected for bulk diffusion in Al, about $0.8\Omega$ and $0.9\Omega$, respectively \cite{Beke} ($\Omega$ is the atomic volume). For the present analysis, the influence of elastic strains on diffusion barriers will rigorously be calculated for the Co atoms in the Al lattice and these values will be used to further estimate Co GB diffusion.

The minimum energy path and the diffusion barrier for a vacancy-mediated nearest neighbor jump of Co in a face centered cubic (fcc) Al matrix is obtained employing the climbing image nudged elastic band (CI-NEB) method\cite{CINEB1,CINEB2}. The diffusion barrier is calculated as the difference between the energies of the saddle point and the initial equilibrium state along the minimum energy path in the configuration space along the [110] direction connecting the initial and the final transition state (see Fig.~\ref*{fig:diffbarrier}b). For the present case, the saddle point lies midway along this minimum energy path. The strain dependence of the diffusion barrier is obtained by calculating the latter in a hydrostatically strained Al matrix for different values of the lattice parameter in order to address the experimentally-relevant effect of hydrostatic pressure on diffusion. 

\subsection{Computational details}

\noindent We have used $2\times2\times2$ and $3\times3\times3$ fcc-supercells (32 and 108 atoms respectively) with one Co atom and one vacancy on a nearest-neighbor site of Co to compute the minimum energy path employing the CI-NEB method\cite{CINEB1,CINEB2} as implemented in the VTST package\cite{vtst}. Both the supercells were generated using the calculated equilibrium lattice parameter (4.04 $\text{\AA}$) of pure Al. The total energy calculations were performed employing the projector augmented wave (PAW) method as implemented in the VASP\cite{Kresse1993,Kresse1996} package with energetics based on density functional theory (DFT). The generalized gradient approximation (GGA) within the Perdew-Burke-Ernzerhof (PBE) parameterization scheme\cite{PBE} was used to describe the electronic exchange and correlation effects. The integration over the Brillouin zone was performed using the Monkhorst-Pack\cite{Monkhorst1976} scheme with an $12\times12\times12$ (for 32 atoms) and $8\times8\times8$ (for 108 atoms) reciprocal-space $k$-mesh centered around the $\Gamma$ point and utilizing the Methfessel-Paxton scheme\cite{Methfessel} with a thermal smearing width of 0.15 eV to account for the smoothening of partial occupancies of the electronic states. The plane-wave energy cutoff was set to 400 eV. An energy of $10^{-6}$ eV was used as a convergence criteria for the self-consistent electronic loop. A tolerance value of 0.05 eV/\AA\ was chosen for the force components along and perpendicular to the tangent to the reaction path. A total of 5 images/configurations (in addition to the initial and the final state) were considered to generate the minimum energy path in the configuration space.

The strain dependence of the diffusion barrier was obtained by performing the total energy calculations for 13 equally spaced hydrostatic strain ($\epsilon$) values ranging between $\pm 1.5\%$ (-1.5, -1.25, -1, \ldots, 1, 1.25, 1.5) where $\epsilon=(a-a_{\rm eq})/a_{\rm eq}$, $a_{\rm eq}$ being the computed equilibrium lattice parameter of pure Al (4.04 $\text{\AA}$) as obtained within GGA. With this definition of $\epsilon$, the negative and positive values correspond to compressive and dilatational strains respectively. The internal atomic coordinates were fully relaxed during the total energy calculations maintaining the strained volume. 

\section{Results}

\begin{figure}[ht]
\centering
\includegraphics[width=8cm]{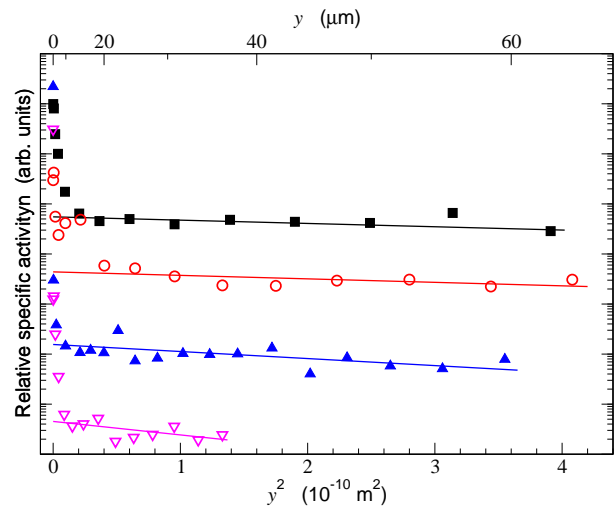}
\caption{Examples of the penetration profiles for $^{57}$Co tracer diffusion at room temperature in the AA5024 alloy in the following states: as-cast (squares), ECAP-processed (circles), ECAP-processed and annealed at 623~K for 1~h (triangles up) and at 773~K for 20~h (triangles down). \label{fig:prof}}
\end{figure}

\noindent Examples of the penetration profiles of $^{57}$Co tracer diffusion in the AA5024 alloy for the different states are shown in Fig.~\ref{fig:prof}. The first near-surface points are affected by the grinding procedure and omitted from the subsequent analysis. The bulk diffusion coefficient of Co in pure Al at 320~K is about $10^{-30}$~m$^2$/s as it follows from the Arrhenius parameters reported by Hood et al. \cite{Cobulk}. Similar values -- at least, not extremely enhanced -- can safely be assumed for Co volume diffusion in the Al-alloy, too, taking into account the relatively low amount of the alloying components and the typical values of the solute enhancement factors \cite{LC}. 

The exact values of the volume diffusion coefficients are not relevant, since the effective GB width, $s\cdot \delta$, is significantly larger than the effective diffusion length in the bulk (condition for C-type diffusion\cite{PD}), i.e. the crucial parameter

\begin{equation}
 \alpha = \frac{s \cdot \delta}{2\sqrt{D_\vv t}}, \label{eq:alpha}
\end{equation}

\noindent is significantly larger than unity. In this expression, $D_\vv$ is the volume diffusion coefficient of Co in Al, and $t$ is the diffusion time. The GB width $\delta$ was measured for fcc metals to be about $0.5$~nm \cite{D, Dasha} and the segregation factor for Co at Al grain boundaries $s$ can be expected to be $\ge 1$. Note that simple estimates prove that $\alpha > 100$ already for $s=1$.

The determined diffusion coefficients, $D_\gb$, of Co diffusion in the Al-based AA5024 alloy in different investigated states are listed in Table~\ref{tab:par}. 

For a reference, Co GB diffusion in pure Al was measured, too (Table~\ref{tab:par}). For this aim, after polishing, the polycrystalline Al sample was annealed at 823~K for several hours and slowly cooled to room temperature to remove all effects of preparation procedure and assure nearly equilibrium conditions for grain boundary segregation of residual impurities.

\section{Discussion}

\noindent In the present work, the tracer diffusion coefficients of Co atoms in Al-based AA5024 alloy and pure Al are measured for different material states (Table~\ref{tab:par}). Several ~~striking features have to be highlighted.

\begin{enumerate}
 \item Co GB diffusion is faster in well-annealed coarse-grained pure Al with respect to the as-cast Al-based AA5024 alloy;
 
 \item Co GB diffusion reveals almost similar rates in as-cast and ECAP-processed states of the Al-based AA5024 alloy;
 
 \item There is an unexpected trend with a strongly non-monotonous dependence of the Co GB diffusion coefficients in the ultrafine grained Al alloy on the annealing parameters, in fact they first decrease and then increase again reaching almost the level which is characteristic for the Co GB diffusion rates in as-cast or ECAP-processes states.

\end{enumerate}

\subsection{Effect of segregation on GB diffusion}

\noindent The feature $1$ may be explained by the alloying effect on GB diffusion, since it follows a general trend observed experimentally that the purer is the material, the higher is the GB diffusion rate; as it was found, e.g., for GB self-diffusion in pure Cu \cite{SH} or Ni \cite{G2, Dasha} or solute (Ni) GB diffusion in pure Cu \cite{Tokey}. It is known that Mg does segregate to grain boundaries in this alloy after annealing at slightly elevated temperatures \cite{Yulia} as it does in other Al--Mg alloys \cite{Xavier}.

In the present case of the ECAP-processing at 573 K and post-annealing at higher temperatures, the chemical analysis did not reveal any measurable segregation at most of the GBs (about 80\%) in the limits of the uncertainty of the EDX measurements. In some of the GBs, a slight increase of the local Mg concentration by about 0.5\% could be detected. Still, a significant increase of the Mg concentration was observed at triple junctions. 

Mg segregation was observed in the as-cast AA5024 alloy \cite{Yulia}, however, Co GB diffusion proceeds with similar rates in as-cast and annealed states with quite different levels of the Mg segregation. Thus it is not Mg segregation which mainly affects Co GB diffusion in the AA5024 alloy and some minority components in the alloy are probably responsible for the diffusion retardation in the alloy with respect to the pure Al grain boundaries. Moreover, GB reconstruction in the AA5025 alloy (following, e.g., the GB complexion theory \cite{Dillon}) might be responsible, too. 

\subsection{Effect of SPD-processing on GB diffusion}

\noindent The present results substantiate that the severe plastic deformation of the Al-based alloy, alternatively to the case of ECAP-processing of Ni or Ti, does not produce a deformation-modified state of the grain boundaries which would be characterized by enhanced diffusion coefficients (feature $2$). A decrease of the GB diffusion coefficients as a result of the ECAP-processing was already observed for the case of Co diffusion in ultrafine grained Ti. However this represents a very special case in view of the interstitial diffusion mechanism of Co atoms in the hexagonal close-packed (hcp) lattice of $\alpha$-Ti \cite{Jochen}. On the other hand, vacancy-mediated GB diffusion of Ag atoms was extremely enhanced in ECAP-processed $\alpha$-Ti \cite{Jochen}. These facts correspond to the generation of excess free volume at deformation-modified GBs which serve as traps for interstitially diffusing Co and as vehicles for substitutionally diffusing Ag atoms. Since the substitutional diffusion mechanism holds definitely for Co in Al \cite{Cobulk}, one may safely assume substitutional solubility and a vacancy-mediated diffusion mechanism for Co atoms in grain boundaries of Al and Al-alloys, too.

The absence of an enhancement of Co GB diffusion after SPD-processing of Al-based alloy is most probably related to a relatively high homologous temperature of the diffusion measurements, $T/T_m$, and need a clarification (here $T_m$ is the corresponding melting temperature).

\subsection{Relaxation of the deformation-modified state}

\noindent Nazarov and co-workers \cite{Nazarov} described a deformation-modified (`non-equilibrium') state of grain boundaries in terms of arrays of extrinsic grain boundary dislocations. These defects were proposed to relax by dislocation climb and annihilate at elevated temperatures. The corresponding relaxation time, $\tau$, was estimated as \cite{N2}

\begin{equation}
 \tau = \frac{k_{\rm B}T d^3}{A\delta D_\gb^{sd} G \Omega}
\end{equation}

\noindent where $k_{\rm B}$ and $T$ are the Boltzmann constant and the absolute temperature, respectively, $d$ is the average grain size, $\delta$ the GB width, $D_\gb^{sd}$ the self-diffusion coefficient for general relaxed high-angle grain boundaries, $G$ the shear modulus, $\Omega$ the atomic volume, and $A$ is a geometrical factor. The value of $A$ was suggested to lie between 100 and 500, depending on the specific disclination model of the ultrafine grained materials used \cite{N2} and it was shown that this value gives reasonable estimates for SPD-processed Ni \cite{Gerrit}. As stated above, the diffusional GB width $\delta \approx 0.5$~nm \cite{Dasha}. The GB self-diffusion coefficient in Al, $D_\gb^{sd}$, is estimated at about $2.8 \times 10^{-16}$~m$^2$/s at 320~K based on the Arrhenius parameters reported in Ref.~\onlinecite{Hoesler}. Using the material parameters of Al ($G$ = 26~GPa and $\Omega = 17.2$~\AA$^3$), the relaxation time of the deformation-modified state of the grain boundaries in Al is on the order of seconds at the measurement temperature of 320 K. Taking into account a finite time, required for a deformed billet to be cooled down to the room temperature from the deformation temperature of 573~K, one may safely conclude that a hypothetical enhancement of the diffusion rates due to a deformation-modified state of GBs, if it does exist during deformation, relaxes and cannot be measured in the present post-mortem experiments.

This conclusion fully agrees with the present results on GB diffusion of Co after ECAP-processing.

\subsection{Effect of crystal dislocations}

\noindent In addition to the excess GB dislocations and the GB modifications, SPD processing is known to induce a high density of crystal dislocations which may approach the values on the order of  $10^{15}$~m$^{-2}$ for Ni or Cu \cite{Schafler}. Dislocations in metals are known to represent generally further short-circuits for enhanced diffusion \cite{PD, Mehrer} and may modify diffusion transport during measurements of GB diffusion \cite{Geise}. Note that deformation-induced vacancies annihilate below room temperature in plastically deformed Al \cite{ZB} and cannot affect the present diffusion measurements.

At least two kinds of models have to be analyzed:

{\bf Dislocation model 1}. Since the C-type conditions are fulfilled, the crystal dislocations represent short-circuits and provide a diffusion enhancement with respect to bulk diffusion in addition to grain boundaries. However, the rates of dislocation diffusion correspond typically to about 1/10$^{\rm th}$ of the diffusion rate along general high-angle GBs \cite{PD, Sommer} and, thus, the dislocation diffusion contribution should not be observed in the present experiments apart from several first near-surface points of the concentration profiles in Fig.~\ref{fig:prof}). Although the dislocation density is larger in ECAP-processed material by orders of magnitude, GB diffusion is slower in the ultrafine grained state with respect to that in pure Al. Furthermore, post-deformation annealing treatment even enhances GB diffusion, while a dramatic decrease of the dislocation density is observed and therefore opposite to what would be expected within this model. We conclude that Model 1 is not applicable to the present case.

{\bf Dislocation model 2}. The crystalline dislocations are attached not only to the free surface of the sample, but they cross the GBs, too. In the latter case, these dislocations represent the paths for enhanced leakage of the tracer atoms from grain boundaries. In the present case with the absence of bulk diffusion, the GB diffusion problem with out-diffusion via dislocation corresponds to the so-called B-C-type regime after Divinski et al. \cite{DHK1, DHK2} or D$_1$-type regime after Klinger and Rabkin \cite{Geise, RK}. The tracer leakage from GBs is controlled by the parameter $\alpha^*$, which is (compare with Eq.~(\ref{eq:alpha}))

\begin{equation}
 \alpha^* = \frac{s \delta}{2A_\dd \rho_\dd \sqrt{D_\dd t}}\label{eq:a*}.
\end{equation}

\noindent Here $g_\dd = A_\dd \rho_\dd$ is the volume fraction of sites belonging to the dislocation pipes, with $A_\dd$ being the cross-section of the dislocation pipe and $\rho_\dd$  the dislocation density. In the case of $\alpha^* > 1$, the C-type regime continues to hold and Eq.~(\ref{eq:Dgb}) has to be applied to determine the corresponding GB diffusion coefficients. The situation changes if $\alpha^* < 0.1$ (and $\alpha$ remains to be large, $\alpha > 1$) \cite{Geise}, in which case the concentration profiles have to be analyzed in the coordinates of the logarithm of concentration vs. the penetration depth $y$ (i.e. for the quasi B-type conditions) and the triple product

\begin{equation}
P = s^* \cdot \delta \cdot D_\gb = 1.128 \sqrt{\frac{D_{\eff}}{t}} \left( \frac{\partial \ln \bar{c}}{\partial y} \right)^{-2} \label{eq:P}
\end{equation}

\noindent has to be determined instead of the diffusion coefficient $D_\gb$ \cite{Geise}. Here,

\begin{equation}
 D_{\eff} = g^2_\dd D_\dd,
\end{equation}

\noindent is the effective diffusivity, which describes the tracer leakage from GBs into the crystal volume by dislocation pipe diffusion. In Eq.~(\ref{eq:P}), $s^*$ is the segregation coefficient for the tracer atoms between the grain boundaries and the dislocation pipes which is probably a value of the order of unity. Estimating the diffusion coefficient along the dislocation pipes as $D_\dd = 1/10 D_\gb$ \cite{PD}, the value of the parameter $\alpha^*$ can be determined (Table~\ref{tab:par}). 

An analysis reveals that $\alpha^* > 1$ for most of the conditions and the effect of dislocations can safely be neglected, thereby confirming that Eq.~(\ref{eq:Dgb}) was correctly used. The only result of the diffusion measurements for Co GB diffusion in ECAP-processed Al alloy with the highest dislocation density (without annealing) has to be revisited, since the corresponding value of $\alpha^*$ is less than unity (Table~\ref{tab:par}). From the GB diffusion theory \cite{PD}, it is well known that GB diffusion under conditions of $0.1 < \alpha < 1$ corresponds to a transition regime between the B- and C-types kinetics and the determined diffusion coefficients underestimate the real values \cite{Szabo}. Following the approach suggested in Ref.~\onlinecite{Szabo} and outlined in Ref.~\onlinecite{PD}, a correction factor can be determined and the corrected value of the corresponding diffusion coefficient is listed in Table~\ref{tab:par}.

\subsection{Effect of particles on GB diffusion}
\label{sc:ParticleEffect}

\begin{figure*}[t]
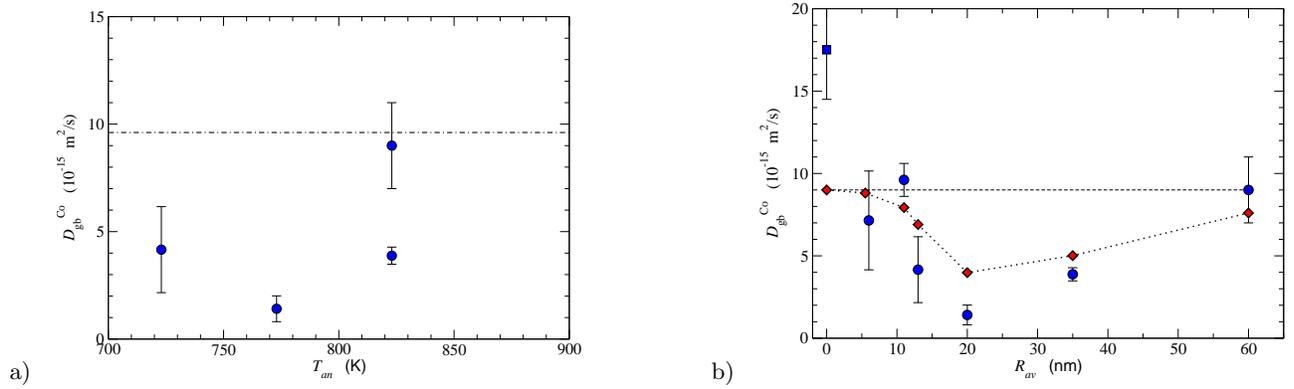

\centering
a)~~ \includegraphics[width=7cm]{figure02a}
\hspace{1.5cm}
b)~~ \includegraphics[width=7cm]{figure02b}

\caption{The dependence of the diffusion coefficient of Co in the ECAP-processed Al alloy on the post-deformation annealing temperature (a) and the particle radius in different states, circles (b). In (a) the GB diffusion coefficient measured for Co after ECAP deformation is shown by the dot-dashed line. In (b) the diffusion coefficient in a particle-free alloy is shown by the dashed line. The diamonds represent the dependence of the diffusion coefficient (normalized with respect to the particle-free case) on the particle size, that includes an implicit dependence on the associated elastic fields produced by the particles as described by the model outlined in  subsection~\ref{sc:model}. The value of $D_\gb^\Co$ in pure Al is shown, too (square). \label{fig:Dgb}}
\end{figure*}

\noindent In Fig.~\ref{fig:Dgb}a the measured diffusion coefficients $D_\gb$ for Co in the ECAP-processed AA5024 alloy are plotted as function of the post-deformation annealing temperature $T_{\an}$ (symbols) and are compared with the value measured in the as-deformed alloy (dot-dashed line). It is obvious that the dependence cannot be attributed to a usual relaxation-like behavior, since annealing at 823~K for long times increases the diffusion coefficients, although annealing treatments at lower temperatures decrease them, Fig.~\ref{fig:Dgb}a. This unusual temperature dependence needs clarification.

\begin{figure*}[t]
a)~\includegraphics[height=4.3cm]{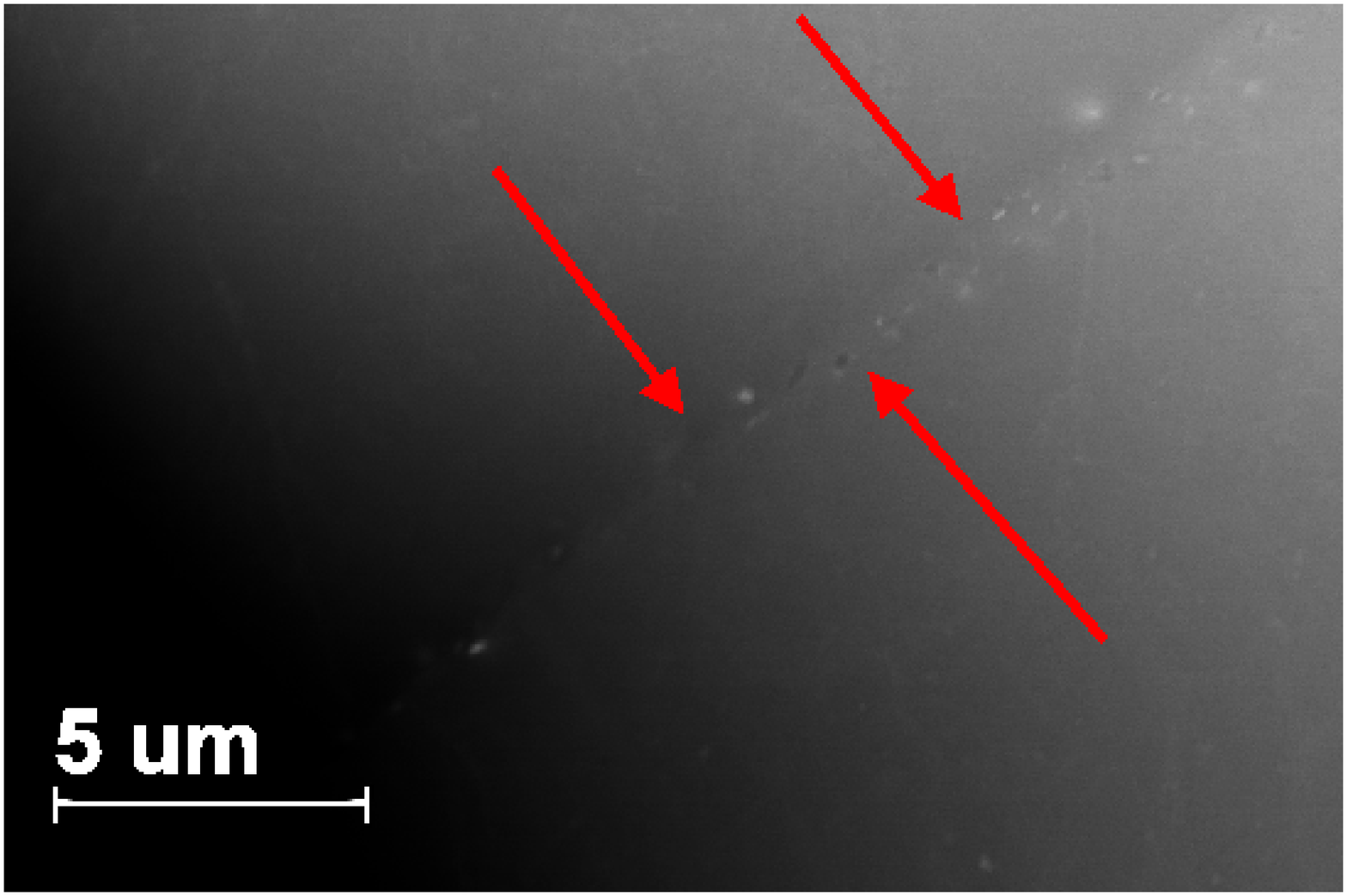}
~~b)~\includegraphics[height=4.3cm]{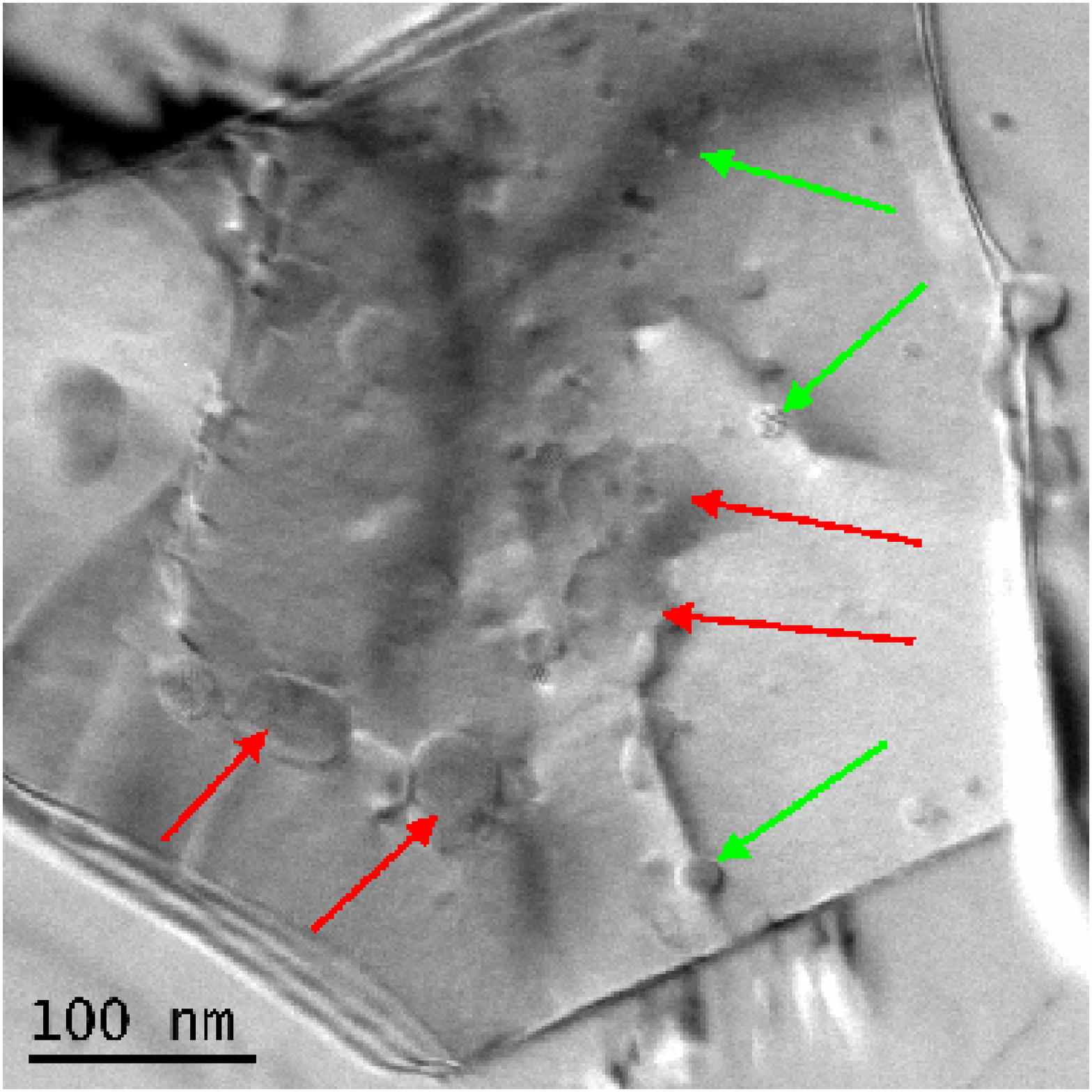}
~~c)~\includegraphics[height=4.3cm]{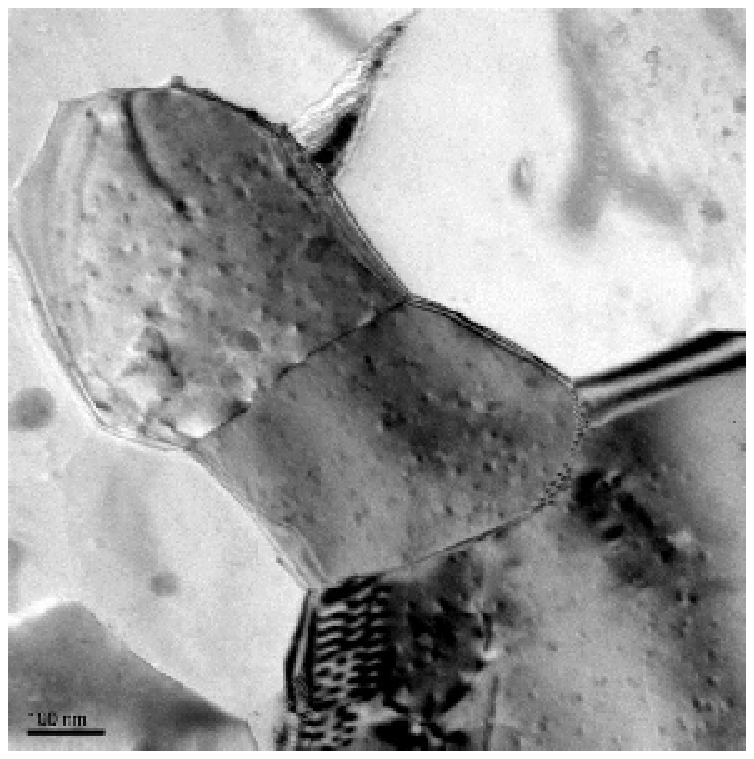}

\caption{\label{fig:TEM} a) HAADF-STEM image of a GB in AA5024 in the as-cast state. The high concentration of Al$_6$Mn precipitates on the GB is indicated by red arrows. b) TEM image of a single grain in AA5024 after ECAP at 573~K. Al$_6$Mn and  Al$_3$Sc precipitates are indicated by red and green arrows, respectively. c) Distribution of Al$_3$Sc-based precipitates in grains and around interfaces in AA5024 after ECAP at 573~K and annealing at 723~K.}
\end{figure*}

Figures \ref{fig:TEM}b and c show a typical contrast (including the appearance of Moir\'{e} patterns) at GBs in different states. It is obvious that the interfaces are straight and flat and appear as relaxed interfaces. This finding agrees perfectly with the results of GB diffusion measurements, which substantiate an absence of deformation-enhanced GB diffusion in the Al-based AA5024 alloy under consideration.

Al$_6$Mn precipitates are found at interfaces as it is substantiated by the HAADF-STEM image, Fig. \ref{fig:TEM}a. The bright-appearing particles in these conditions are characterized by larger $Z$ numbers of scattering atoms with respect to the matrix (Al) and correspond to Al$_6$Mn precipitates.  

Figure \ref{fig:TEM}a shows that in the as-cast state the Al$_6$Mn precipitates are mainly concentrated near GBs. The average size of these precipitates is about 200 nm with relatively large average distance between them, more than 300~nm. 
While the nano-sized Al$_3$Sc-based particles are typically coherent in the as-deformed AA5024 alloy, the larger Al$_6$Mn particles are incoherent with the matrix. As discussed below, a relatively small contribution of these particles on the measured Co GB diffusion can be expected.

After ECAP processing, the precipitates (both Al$_6$Mn and Al$_3$Sc) are more uniformly distributed in the grain interiors,  Fig.~\ref{fig:TEM}b. The Al$_6$Mn particles are of a size of about 70 nm and are predominantly located at dislocations and sub-boundaries. The number fraction of particles near the GBs in the ECAP-processed sample is estimated to be about $10^2$ particles per $\mu$m$^3$, which is lower than the total volume fraction (about $10^{4}$ $\mu$m$^{-3}$). Thus the particles are mainly distributed in the grain interiors, keeping the GBs relatively free. The average distance between the precipitates in the deformed states is about 90~nm.

As a result of post-deformation annealing the particles grow and the Al$_3$Sc precipitates approach an average size of 40 nm and more (Table \ref{tab:par}). It is important that whereas Al$_3$Sc precipitates are fully coherent in the ECAP-processed state (their average size is about 11 nm), they become semi-coherent after annealing treatment as a result of growth, and finally loose their coherency after annealing at high temperatures.

The precipitate/matrix misfit can be determined as $(a_\pp - a_\mm)/a_\mm$, where $a_\m$ and $a_\pp$ are the lattice parameters of the matrix and a precipitate, respectively. Using DFT, we have determined a misfit between the Al$_3$Sc precipitates and pure Al matrix of approx.~1.5\% at $T=0$K, which is decreasing with temperature, in agreement with the experimental value \cite{marq} of 1.33\% at room temperature. The critical precipitate size, corresponding to the appearance of misfit dislocations, is then $a_\mm^2/[2(a_\pp - a_\mm)]$ = 15.2 nm, where 2 corresponds to the (002)-Al Burgers vector. 

However, the AA5024 alloy contains up to 4.5 wt.\% of Mg that increases the lattice parameter of the matrix \cite{murr} thereby decreasing the misfit. Moreover, in the ECAP-processed state the Al$_3$Sc particles contain Zr (and sometimes Ti) \cite{Yulia} that decreases the lattice parameter of the precipitates. We performed an analysis using the data from Ref.~\onlinecite{harada} (see Table~\ref{tab:misfit}) and the results show that the Al$_3$(Sc,Zr) precipitates in the Al--4.5wt.\%Mg matrix start to loose their coherency with the matrix at the sizes of about $20-30$~nm  (depending on the Zr content). The Ti-containing Al$_3$(Sc,Ti) precipitates will not be analyzed further, since their volume fraction is found to be low \cite{Yulia}.

\begin{table}[h]
\begin{center}
\footnotesize
\caption{\label{tab:misfit} The summary of critical sizes for Al$_3$Sc-based precipitates in the case of pure Al and the Al--Mg alloy. The calculations were performed using the data from Ref.~\onlinecite{harada}.}

\begin{tabular}{c|ccc}
\hline
Matrix & Particle & \parbox[c][0.8cm]{1.8cm}{Mismatch (\%)} & \parbox[c][0.8cm]{2cm}{critical particle size (nm)}\\
\hline
 & Al$_3$Sc & 1.33 & 15.2 \\

Al & Al$_3$(Sc,Zr) & 1.14 & 17.7 \\

\vspace{1mm}
& Al$_3$(Sc,Ti) & 0.64 & 31.6 \\

 & Al$_3$Sc & 0.87 & 23.4 \\

\parbox[c][0.5cm]{2cm}{Al+4.48\%Mg} & Al$_3$(Sc,Zr) & 0.67 & 30.4 \\

& Al$_3$(Sc,Ti) & 0.18 & 113.0 \\
\hline
\end{tabular}
\normalsize
\end{center}
\end{table}

\begin{figure}[ht]
 \includegraphics[width=8cm]{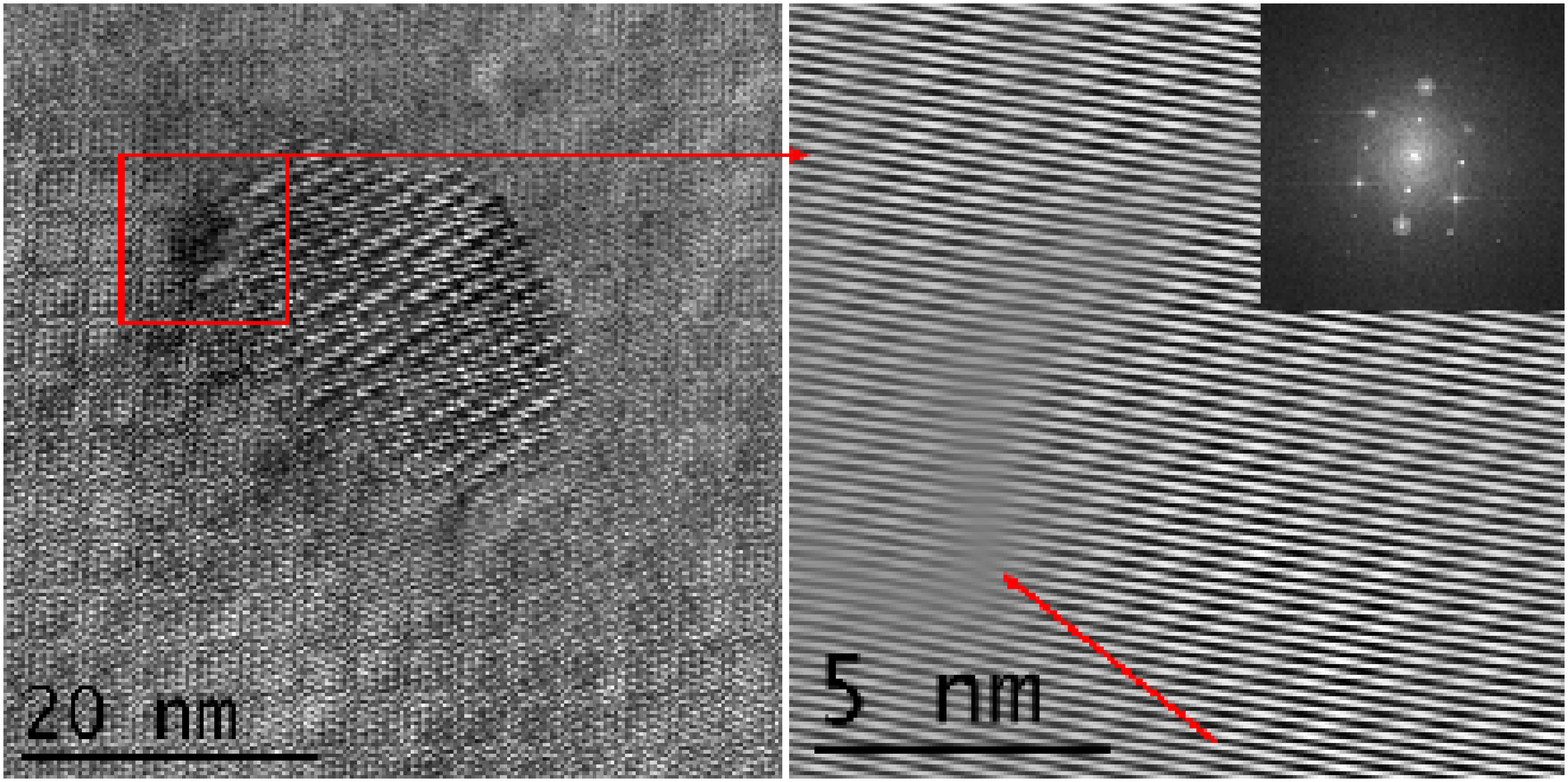}
 
 \caption{Left: HR-TEM image of an Al$_3$(Sc,Zr) precipitate in the AA5024 alloy after ECAP processing at 573~K and subsequent annealing at 673~K for 1~h. Right: FFT-analysis of the precipitate interface in the (002)-Al Bragg maxima. A misfit dislocation is indicated by the arrow.\label{fig:misfit}}
\end{figure}

Figure~\ref{fig:misfit} shows a dislocation analysis of the interface structure for an Al$_3$Sc precipitate. The size of the particle is about 22~nm and the appearance of misfit dislocations has to be expected according to the analysis (Table~\ref{tab:misfit}). The precipitate is not perfectly spherical, which correlates with a relatively low content of Zr atoms\cite{Yulia}. A fast Fourier transformation (FFT) image made in the  (002)-Al Bragg maxima indeed verifies the existence of a misfit dislocation between the matrix and the particles (red arrow in Fig.~\ref{fig:misfit}). 

In our previous work on Ni and Ti GB diffusion in NiTi alloy, a retardation of the GB diffusion rate was measured at high temperatures and was related to the precipitation of Ni$_4$Ti$_3$ particles at the interfaces \cite{Dandan}. In the present work, we will analyze the effect of precipitates on GB diffusion in the Al-based severely deformed alloy.

Figure~\ref{fig:Dgb}b re-plots the measured diffusion coefficients $D_\gb$ as a function of the averaged radius of the Al$_3$Sc-based precipitates, $R_{\av}$. A strongly non-monotonous dependence is seen where the Co GB diffusion coefficient first decreases with increasing radius, reaching a critical value at about $R_{\av} \approx 20-30$~nm, followed by an increase and approaching the value of $D_\gb$ for small precipitates. Note that this is exactly the critical size for Al$_3$Sc and Al$_3$(Sc,Zr) precipitates when appearance of misfit dislocations and a gradual loose of the precipitate--matrix coherency have to be expected (Table~\ref{tab:misfit}).

In following subsections, we will analyze the effect of these precipitates on the GB diffusion.


\subsection{Elastic field around a precipitate}

\noindent The stress field caused by the Al$_3$Sc precipitates in the matrix was calculated according to Eshelby's solutions for spherical inclusions in an infinite isotropic and elastic body \cite{Eshelby}. The radial dependencies for the strain, $\epsilon_r$, and stress, $\sigma_r$, fields are,

\begin{equation}
 \epsilon_r = \frac{1+\nu}{3(1-\nu)} \left( \frac{a}{r} \right)^3 \left( \Delta a + \Delta \gamma \Delta T \right)\label{eq:strain}
\end{equation}

\begin{equation}
 \sigma_r = E \epsilon_r \quad .\label{eq:stress}
\end{equation}

\noindent Here, $\nu$ is the Poisson ratio, $a$ is the radius of a precipitate, $r$ the radial distance from the precipitate center, $\Delta a$ the lattice misfit between the precipitate and the matrix, $E$ the Young modulus of the matrix, and $\Delta \gamma$ is the difference of the thermal expansion coefficients between the precipitate and the matrix. We have taken into account that the precipitates are formed and relaxed during hot-deformation or annealing, while diffusion measurements are performed at nearly room temperature. 
The value of $\Delta \gamma$ is taken from DFT-based calculations \cite{Ankit} which were performed for both, pure Al and Al$_3$Sc including vibrational and electronic contributions, and it was further shown that anharmonicity hardly affects the values.

The elastic strains (compressive) are maximum at the particle surface and decay with increased distance from the particle. Typically, the strain becomes small (less than $-0.1$\%) at the double particle radius. Elastic stresses behave qualitatively similarly, Eq.~(\ref{eq:stress}). Note that the yield stress of the AA5024 alloy is 530~GPa \cite{Mog3}.

In the case of vacancy-mediated diffusion, the diffusion coefficient of a solute, $D$, is generally,

\begin{equation}
 D = \frac{1}{6} f \lambda^2 \nu_0 \exp \left( - \frac{\Delta G^f + \Delta G^b + \Delta G^m}{k_{\rm B}T} \right).  \label{eq:diff-rate}
\end{equation}

\noindent Here, $\lambda$ is the jump distance, $\nu_0$ the attempt (Debye) frequency, $f$ the correlation factor (temperature-dependent for solute diffusion), and $\Delta G^f$, $\Delta G^b$ and $\Delta G^m$ are the free energies of vacancy formation, solute--vacancy binding, and of vacancy migration respectively. Note that the present tracer diffusion experiments are performed under nearly equilibrium conditions, since the deformation-induced vacancies in deformed Al matrix relax already at room temperature \cite{ZB}.

\begin{figure*}[t]
	\includegraphics[width=7cm]{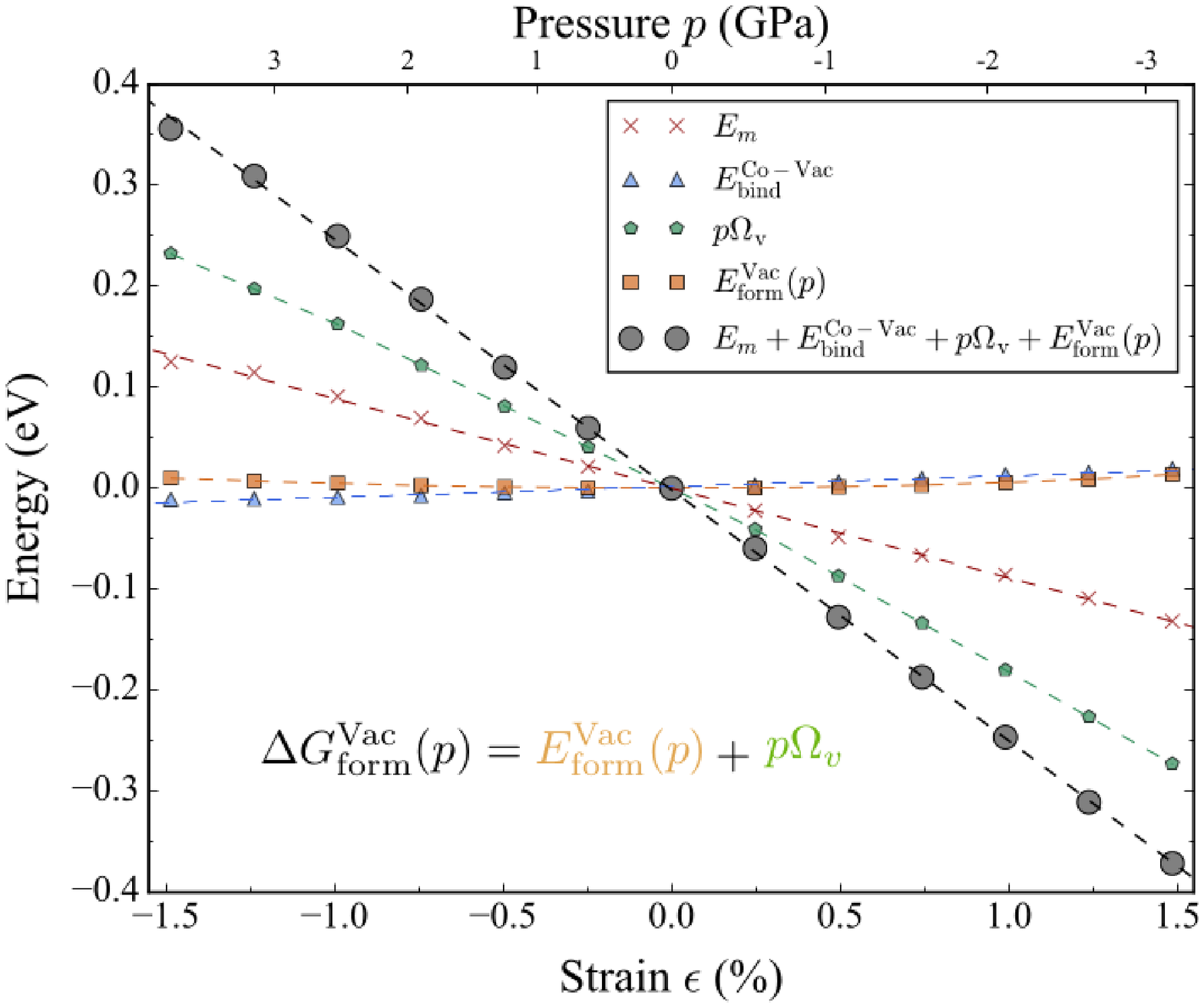}
	\hspace{1.5cm}
	\includegraphics[width=8cm]{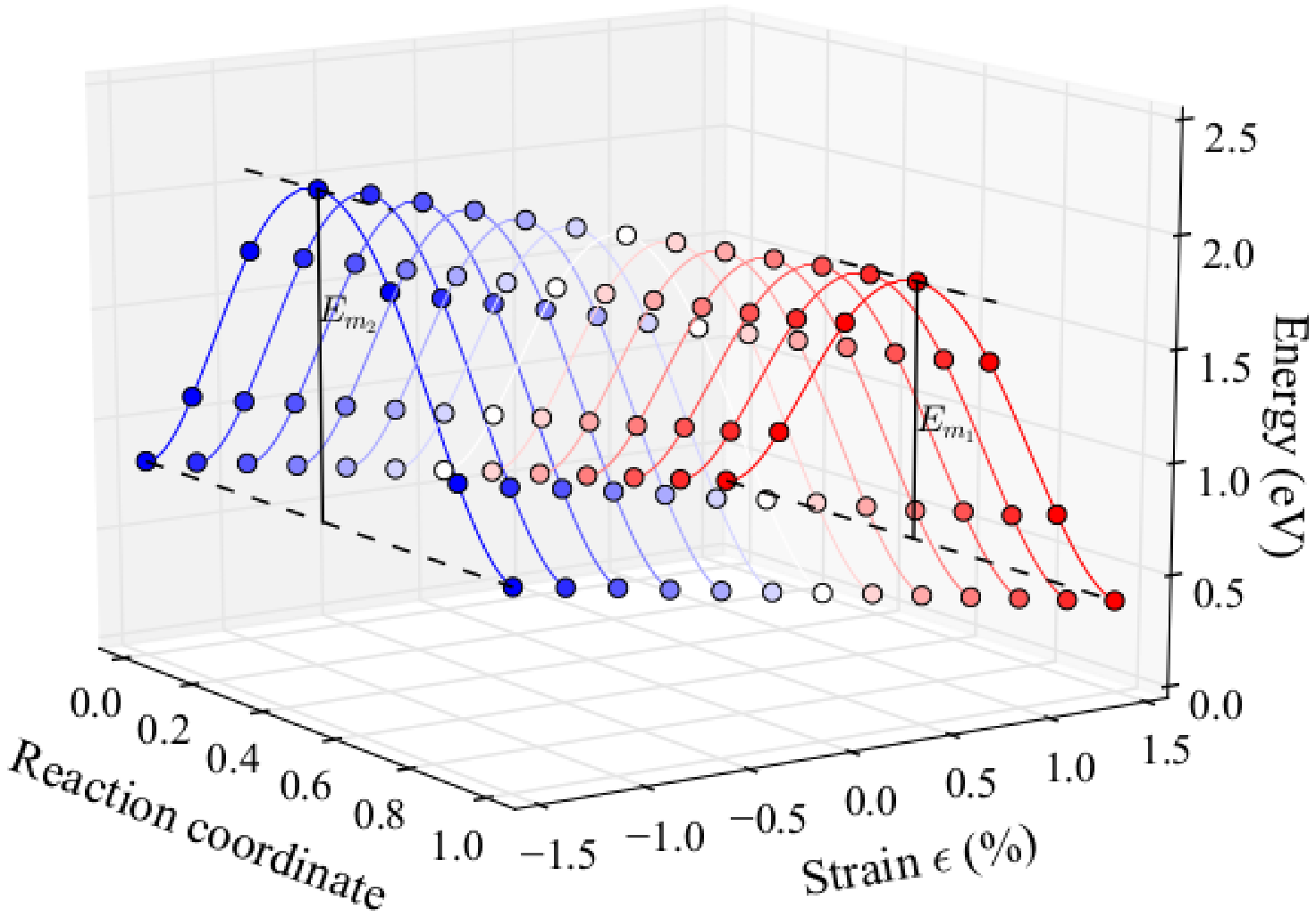}
	\caption{(a) Strain dependence of the vacancy formation energy $E^{\rm Vac}_{\rm form}$, the Co-vacancy binding energy $E^{\rm Co-Vac}_{\rm bind}$, and the Co migration energy $E_{\rm m}$ for vacancy mediated nearest neighbor jumps in pure Al. In addition the pressure-volume work $p \Omega_f$ is provided, which also enters the sum of all energy contributions. (b) 3D representation of the variation of the energy profile (for Co atom diffusing into the nearest neighbor vacancy) along the minimum energy path for different strain rates. The respective intial state corresponds to the Gibbs energy of a vacancy formation next to a Co atom. \label{fig:diffbarrier}}
\end{figure*}

In the following we will neglect the effect of strain on the solute--vacancy binding energy as well as on the chemical part of the vacancy formation energy. 
Direct DFT-based calculations, see below, verify this assumption.
The elastic part of the vacancy formation energy depends on the hydrostatic pressure $p$ induced by precipitates via a $p\Omega_\vv = \frac{1}{3}\sigma_{ii}\Omega_\vv$ term. Here, $\Omega_\vv$ is the vacancy formation volume. It is known that the migration volume in the fcc lattice is small \cite{Mehrer} and we will compute the effect of pressure on the formation volume.

In the present estimates, the correlation factor $f$,  which can in principle be calculated in terms of the five-frequency model \cite{LCL} for the fcc lattice, is considered as a constant (at the given temperature of the diffusion measurements). Moreover, the effect of pressure on the total term $f \lambda^2 \nu_0$ is neglected\cite{Mehrer}. This is  a reasonable asumption, since the strain dependencies in the exponent are more significant than in the prefactor.

Thus, the change of the GB diffusion coefficient of Co atoms due to applied elastic stresses is estimated as

\begin{equation}
 \frac{D_\gb^\Co(\sigma)}{D_\gb^\Co(0)} = \exp \left( \frac{E_m(0) - E_m(\sigma)}{k_{\rm B}T} \right) \exp \left(- \frac{\frac{1}{3} \sigma_{ii} \Omega_\vv}{k_{\rm B}T} \right).\label{eq:dif-ret}
\end{equation}

As an estimate, the value of the vacancy formation volume measured for Zn GB diffusion in Al \cite{Beke}, $\Omega_\vv = 0.8\Omega$, can be used for Co GB diffusion in Al. Now, we need to determine  the strain dependence of $E_m(\sigma)$.

\subsection{Ab initio energetics of Co diffusion in Al}

\noindent Figure~\ref{fig:diffbarrier} shows the ab initio calculated energy contributions entering the exponent in Eq.~(\ref{eq:diff-rate}). They have initially been determined as Helmholtz energies for hydrostatic strain values up to $\pm 1.5\%$. The maximum elastic strain of $1.5\%$ corresponds to the elastic misfit between the pure Al matrix and Al$_3$Sc particles, determined in subsection~\ref{sc:ParticleEffect}. 
By performing the calculations for $2\times2\times2$ and $3\times3\times3$ supercells we confirmed that finite size errors at the maximum strain are below 5 \%. The convergence is determined by the point defect interaction energies as a function of distance in the fcc Al matrix: While the Co-Co interaction energy is 155 meV in a nearest-neighbor configuration, it is already in the 5$^{\rm th}$ nearest neighbor shell decreased to 10\% of this value. The results of $3\times3\times3$ supercells, plotted in Fig.~\ref{fig:diffbarrier} can therefore be considered as sufficiently accurate.

We observe that the vacancy formation energy $E^{\rm Vac}_{\rm form}$, the Co-vacancy binding energy $E^{\rm Co-Vac}_{\rm bind}$, and the Co migration energy $E_{\rm m}$ depend all almost linearly on the applied strain. The term $E^{\rm Vac}_{\rm form}(p)$ has been determined from defect and defect-free structures that are evaluated at the same pressure, using pure fcc Al  as a reference system for the pressure (upper x axis) corresponding to a certain strain state (lower x axis). 
The Legendre transformation from Helmholtz to Gibbs energies further requires the addition of the pressure-volume work $p \Omega_{\rm f}$ with the vacancy formation volume
\begin{equation}
   \Omega_{\rm f} = V(N-1, p) - \frac{N-1}{N} V(N,p), 
\end{equation}
where $V(N,p)$ is the volume of a supercell with $N$ atoms at pressure $p$. The largely strain-independent term $E^{\rm Vac}_{\rm form}(p)$ justifies the experimental approach, to identify $\Delta G_f$ in Eq.~(\ref{eq:diff-rate}) with $p \Omega_{\rm v}$. 

The migration barrier energy cannot be directly obtained from experiment and we therefore use the result from DFT calculations for the present evaluation. The full diffusion profile for the different strain states is shown in Figure~\ref{fig:diffbarrier}b. The respective initial state corresponds to the Gibbs energy of a vacancy formation next to a Co atom as given by the sum $E^{\rm Vac}_{\rm form}(p)+p \Omega_{\rm f}+E^{\rm Co-Vac}_{\rm bind}$. 
In this configuration, Co still has a higher coordination (11-fold) than in the saddle point configuration. This could explain why the latter is comparatively more sensitive to the compressing/stretching (compressive/dilatational strain) of the lattice (see Figure~\ref{fig:diffbarrier}b). Nevertheless, also the migration barrier of Co in Al follows to a good approximation a linear dependence on the applied strain, for which the expression 

\begin{equation}
 E_m = ( -0.09 \epsilon + 1.31 ) {\rm ~eV} \label{eq:str}
\end{equation}

\noindent  will be used in subsequent estimates.


The calculations further reveal that the maximum change in the binding energy due to the strain is 0.015~eV if $|\epsilon |\le 1.5\%$. This is by an order of magnitude smaller than the change of the migration barrier (0.12~eV). Hence, it is safe to ignore this contribution.

\subsection{Diffusion coefficient}

\noindent To estimate the influence of the particles on the net diffusion flux along GBs in the AA5024 alloy, we use the modified Maxwell-Garnett equation~\cite{Kalnins, BM}, 
\begin{multline}
\frac{D^{\eff}}{D_\gb} = \frac{k}{f_\pp+k(1-f_\pp)} \times \\
\frac{k(1-f_\pp) + (1+f_\pp)\frac{D_{\pp}}{D_\gb}}{k(1+f_\pp) + (1-f_\pp)\frac{D_{\pp}}{D_\gb}}.\label{eq:MG}
\end{multline}

\noindent Here, $f_\pp$ is the volume fraction of precipitates which are characterized by the diffusion coefficient $D_{\pp}$ and the factor $k$ takes into account probable segregation of Co atoms to the Al/Al$_3$Sc (or Al/Al$_3$(Sc,Zr)) interfaces within Al grain boundaries and the equation is written for the 2D case of GB diffusion. 

\begin{figure}[t]
 \includegraphics[width=7cm]{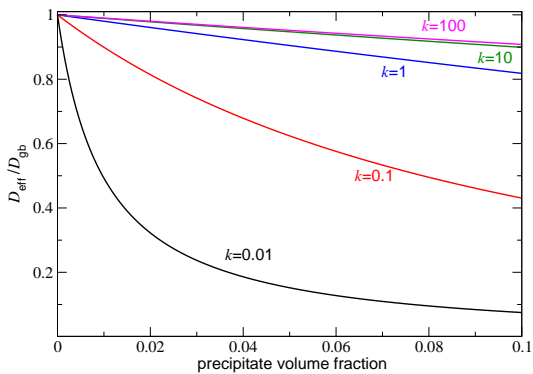}
 \caption{Diffusion retardation in precipitated GBs as a function of the precipitate volume fraction. The dependencies were calculated according to Eq.~(\ref{eq:MG}) for different values of the parameter $k$ (see text). \label{fig:MG}}
\end{figure}

The effect of the precipitate volume fraction on the effective diffusion coefficient is shown in Fig.~\ref{fig:MG} for different values of $k$ and the limiting case $D_\pp = 0$. It is important to recognize that in the original papers, Refs.~\onlinecite{Kalnins, BM}, effective diffusion in a homogeneous media is analyzed with an application to a general case of GB diffusion (in a $d$-dimensional solid \cite{Kalnins}). Then, the segregation was introduced as solute enrichment in the grain boundaries with respect to the crystalline bulk. 

Figure~\ref{fig:MG} shows that even in the case of a low (about 1\%) fraction of precipitates in which diffusion is forbidden, $D_\pp=0$, a strong segregation of tracer atoms to the GB/precipitate interfaces ($k=0.01$) would slow down the GB diffusion rate by a factor of two. This finding already agrees with the experimental observations in the present measurements (Fig.~\ref{fig:Dgb}b). Note that for the critical size of the Al$_3$Sc precipitates of about 20~nm, Table~\ref{tab:par}, and the average distance between them about 200~nm (the measured precipitate density of $10^2$~$\mu$m$^{-1}$ in the near GB region), the particle fraction is estimated at about 1\% in the case of 2D GB diffusion.

\subsection{General model of the particle effect on GB diffusion} 
\label{sc:model}

\noindent In this subsection, we will summarize the obtained results and describe a general model of the diffusion retardation in the precipitated GBs. 

Grain boundaries in ECAP-processed AA5024 alloys contain some fraction of small, size of about $11$~nm, coherent Al$_3$Sc precipitates. The Al$_6$Mn particles are larger, non-coherent, and sparsely distributed. Their contribution to the diffusion retardation is estimated to be small under the assumption that they block GB diffusion locally without introducing long-range strain fields. Thus, we are focusing on Al$_3$Sc precipitates. Still a slight difference between the Co GB diffusion coefficients in as-cast and ECAP-processed states, Fig.~\ref{fig:Dgb}b, may be induced by the Al$_6$Mn particles, Fig.~\ref{fig:TEM}a. The fraction of Al$_3$Sc particles at grain boundaries in ECAP-processed AA5024 alloy is estimated at about 0.2\%.

The net effect of precipitation on Co GB diffusion was estimated according to the following scheme:

\begin{enumerate}
 \item The elastic strain and stress fields around the coherent precipitates are determined according to Eqs.~(\ref{eq:strain}) and (\ref{eq:stress});

 \item The jump barriers, $E_m(\sigma)$, at the given radial positions are determined, Eq.~(\ref{eq:str});
 
 \item The term $\sigma \Omega_\vv$ is determined from Eq.~(\ref{eq:stress}) and using $\Omega_\vv=0.8\Omega$ for grain boundaries;
 
 \item The resulting variation of the GB diffusion coefficient of Co atoms is estimated by Eq.~(\ref{eq:dif-ret}) as a function of the radial distance from the precipitates. 
\end{enumerate}

\begin{figure}[t]
 \includegraphics[width=8cm]{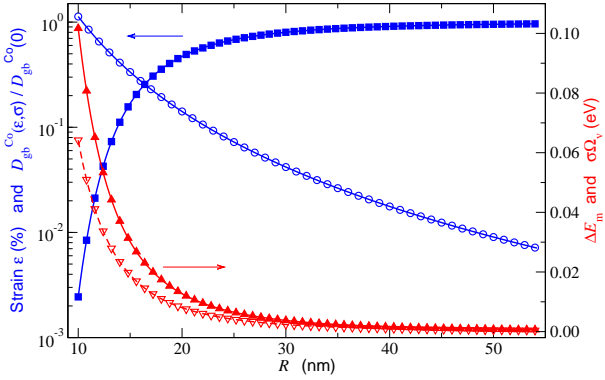}
 \caption{The diffusion retardation, $D_\gb^{\rm Co}(\epsilon)/D_\gb^{\rm Co}(0)$, (blue squares, left ordinate) as a function of radial distance from the particle center $R$ plotted together with the radial dependencies of the elastic strain $\epsilon$ (blue circles, left ordinate), the calculated change of the migration barrier $\Delta E_m$ (red triangles up, right ordinate), and the pressure term $\sigma\cdot \Omega_\vv$ (red triangles down, right ordinate). The 20~nm large coherent Al$_3$Sc particle is considered. \label{fig:D-model}}
\end{figure}

Figure~\ref{fig:D-model} illustrates these calculations for the case of a coherent Al$_3$Sc precipitate of the size of $20$~nm as function of the distance from the precipitate center. The determined elastic strain is plotted by blue circles (left ordinate), the corresponding change of the migration barrier $\Delta E_m(\sigma)$ and the $\sigma \Omega_\vv$ term are shown by triangles up and down, respectively (right ordinate), and the estimate variation of the Co GB diffusion coefficient (normalized on strain-free case) is given by blue squares (left ordinate).

The Co GB diffusion coefficients drop by orders of magnitude at the particle, but approaches its stress-free value at a distance of about three times the radius, Fig.~\ref{fig:D-model}.

Now, Eq.~(\ref{eq:MG}) is used to determine the net effect for the experimentally determined particle fraction $f_\pp$ and the estimated diffusion retardation on the long-range effective Co GB diffusion coefficient (which corresponds to the experimentally measured value). 

Equation ~(\ref{eq:MG}) is used directly for large (non-coherent) particles with zero-strain fields.  

In the case of coherent particles the strain fields around the particles are explicitly taken into account in an iterative way. Since the effective solution Eq.~(\ref{eq:MG}) holds for spherical precipitates \cite{Kalnins}, we introduced four concentric spheres around a particle of the radius $R$ with the radii $1R$, $1.5R$, $2R$, and $3R$ and averaged the effect of strain on the diffusion coefficients for the latter three radii taking it as zero within the precipitate (for the distances $<R$, i.e. considering the particles as impermeable for diffusing atoms). Then Eq.~(\ref{eq:MG}) is sequentially used to estimate the effective diffusivities of the regions within $1.5R$ of the particle size, $2R$ and $3R$ taking the effective diffusivity of the inner sphere as $D_\pp$ for the next one (with a proper rescaling of the particle volume fractions). The last value is used for a final calculation of the effective diffusivity of a GB (here the experimentally determined value of $f_\pp$ was used).

Figure~\ref{fig:Dgb}b represents the final results of the numerical estimates of the diffusion retardation caused by the precipitation. 

There are two undetermined (free) parameters of the model. First, the Co GB diffusion coefficient along a precipitate-free high-angle GB in Al. We used the value measured for the ECAP-processed case and scaled our results accordingly.

The second free parameter is the segregation factor $k$ in Eq.~(\ref{eq:MG}) and the calculations presented in Fig.~\ref{fig:Dgb}b are performed for a moderate segregation, $k=0.1$ (diamonds). A good agreement with the experimental data is seen.

The particle growth induced by the thermal annealing treatments at 723 and 773~K results in an increase of the elastic fields around them, keeping the particles still coherent, and the effective diffusion coefficient drops (Fig.~\ref{fig:Dgb}b). It is the state in which the GB diffusion coefficient of Co atoms is slowest and amounts to about 50\% of the value in the ECAP-processed state. 

As a result of annealing at a higher temperatures, the precipitates grow above the critical size and loose their coherency. Simultaneously, there is a dissolution of the particles and a decrease of the corresponding fraction. Therefore, the Co GB diffusion coefficient increases approaching the values typical for the precipitate-free AA5024 alloy (Fig.~\ref{fig:Dgb}b).

\section{Conclusions}

Grain boundary diffusion of Co in Al-based AA5024 alloy is found to be slower than in pure Al and is not enhanced by severe plastic deformation. The absence of diffusion enhancement after SPD treatment is explained by a fast relaxation of the deformation-modified state even at room temperature.

The deformation-induced dislocations do not practically affect GB diffusion in the ECAP-processed AA5024 alloy.

A post-deformation annealing treatment of the ultrafine grained AA5024 alloy results in a non-monotonous dependence of the Co GB diffusion rate on the annealing temperature, which first decreases up to a certain critical value beyond which it increases again. This behavior is explained by the evolution of the Al$_3$Sc precipitates in the alloy. 

A model is suggested which takes into account the strain fields around coherent Al$_3$Sc precipitates. The effects of strains on Co GB diffusion are determined by DFT-based calculations. Effective diffusion in the precipitated GBs is described as diffusion in a heterogeneous media with diffusion-blocking precipitates. The importance of solute segregation to the precipitates is highlighted.

\begin{acknowledgments}
Financial support from the Deutsche Forschungsgemeinschaft (DFG) within the priority program SPP-1713 \enquote{chemomechanics} (research projects HI 1300/8-1 and DI 1419/7-1) is gratefully acknowledged.
\end{acknowledgments}



\begin{thebibliography}{99}
\bibitem{V1}R. Valiev, Y. Estrin, Z. Horita, T.G. Langdon, M.J. Zehetbauer, Y.T. Zhu, Producing bulk ultrafine-grained materials by severe plastic deformation, JOM 58 (2006) 33--39. 
\bibitem{V2}R. Valiev, R.K. Islamgaliev, I.V. Alexandrov, Bulk nanostructured materials from severe plastic deformation, Prog. Mater. Sci. 45 (2000) 103--189. 
\bibitem{rev}M.J. Zehetbauer, R. Gr\"ossinger, H. Krenn, M. Krystian, R.Pippan, P. Rogl, T. Waitz, R. W\"urschum, Adv. Eng. Mater. 12 (2010) 692--700.
\bibitem{Mog}A. Mogucheva, D. Tagirova, R. Kaibyshev, Superplasticity in a 5024 aluminium alloy processed by severe plastic deformation, Mater. Sci. Forum 735 (2013) 353--358.
\bibitem{Sauvage}X. Sauvage, G. Wilde, S.V. Divinski, Z. Horita, R.Z. Valiev, Grain boundaries in ultrafine grained materials processed by severe plastic deformation and related phenomena
Mat. Sci. Eng. A 540 (2012) 1--12.
\bibitem{Zehet}D. Setman, M.B. Kerber, E. Schafler, M.J. Zehetbauer, Activation Enthalpies of Deformation-Induced Lattice Defects in Severe Plastic Deformation Nanometals Measured by Differential Scanning Calorimetry, Metall. Mater. Trans. A 41 (2009) 810--815.
\bibitem{EV}Y. Estrin, Y. Vinogradov, Extreme grain refinement by severe plastic deformation: A wealth of challenging science, Acta Mater. 61 (2013) 782--817.
\bibitem{Gerrit}S.V. Divinski, G. Reglitz, H. R\"osner, Y. Estrin, G. Wilde, Self-diffusion in Ni prepared by severe plastic deformation: effect of non-equilibrium grain boundary state, Acta Mater. 59 (2011) 1974-1985.
\bibitem{Jochen}J. Fiebig, S.V. Divinski, H. R\"osner, Y. Estrin, G. Wilde, Diffusion of Ag and Co in ultrafine grained $\alpha$-Ti deformed by equal channel angular pressing, J. Appl. Phys. 110 (2011) 083514.
\bibitem{Nazarov}A.A. Nazarov, A.E. Romanov, R.Z. Valiev, On the structure, stress-fields and energy of non equilibrium grain-boundaries, Acta Metall. Mater. 41 (1993) 1033--1040.
\bibitem{IF}S.V. Divinski, G. Reglitz, I. Golovin, M. Peterlechner, G. Wilde, Effect of heat treatment on diffusion, internal friction, microstructure and mechanical properties of ultrafine grained nickel severely deformed by equal channel angular pressing, Acta Mater. 82 (2015) 11--21.
\bibitem{Nariman}M.M. Abramova, N.A. Enikeev, X. Sauvage, A. Etienne, B. Radiguet, E. Ubyivovk, R.Z. Valiev, Thermal stability and extra-strength of an ultrafine grained stainless steel produced by high pressure torsion, Rev. Adv. Mater. Science 43 (2015) 83--88.
\bibitem{Mehrer}H. Mehrer, Diffusion in Solids: Fundamentals, Methods, Materials, Diffusion-controlled Processes, Springer, Berlin, 2007.
\bibitem{Grabski}M.W. Grabski, R. Korski, Grain boundaries as sinks for dislocations, Philos. Mag. 22 (1970) 707.
\bibitem{Grabski1}R.A. Varin, J.W. Wyrzykowski, W. Lojkowski, M.W. Grabski, Spreading of extrinsic grain boundary dislocations in plastically deformed Aluminium, phys. stat. sol. (a) 46  (1978) 565--569.
\bibitem{Gleiter}Y.H. Pumphrey, H. Gleiter, The annealing of dislocations in high-angle grain boundaries, Philosophical Magazine, 30 (1974) 593--602.
\bibitem{Horita}Z. Horita, D.J. Smith, M. Furukawa, M. Nemoto, R.Z. Valiev, T.G. Langdon, An investigation of grain boundaries in submicrometer-grained Al-Mg solid solution alloys using high-resolution electron microscopy, J. Mater. Res. 11 (1996) 1880--1890.
\bibitem{Baluffi}T.E. Hsieh, R.W. Balluffi, Observations of roughening de-faceting phase-transitions in grain-boundaries, Acta Metall. 37 (1989) 2133--2139.
\bibitem{Divinski}S.V. Divinski, Grain boundary diffusion in severe plastically deformed metals: State of the art and unresolved issues, Diffusion Foundations 5 (2015) 57--73.
\bibitem{Dandan}D. Liu, J. Fiebig, M. Peterlechner, S. Trubel, M. Wegner, Y. Du, Zh. Jin, G. Wilde, S.V. Divinski, Temperature-induced grain boundary transformations as revealed by grain boundary diffusion in B2 NiTi alloy, Intermetallics  61 (2015) 30--37.
\bibitem{Sc}J. Royset, N. Ryum, Scandium in aluminium alloys, Int. Mater. Reviews (2005) 19--44.

\bibitem{Yulia}Yu. Buranova, V. Kulitskiy, M. Peterlechner, A. Mogucheva, R. Kaibyshev, S.V. Divinski, G. Wilde, Al3(Sc,Zr)-based precipitates in Al–Mg alloy: effect of severe deformation, Acta Mater. 124 (2017) 210--224.
\bibitem{seg1}X.-Y. Liu, J. Adams, Grain-boundary segregation in Al--10\% Mg alloys at hot working temperatures, Acta Mater. 46 (10) (1998) 3467--3476.
\bibitem{seg2}X. Sauvage, E. Bobruk, M.Y. Murashkin, Y. Nasedkina, N. Enikeev, R. Valiev, Optimization of electrical conductivity and strength combination by structure design at the nanoscale in Al--Mg--Si alloys, Acta Mater. 98 (2015) 355--366.
\bibitem{Mog2}A. Mogucheva, E. Babich, B. Ovsyannikov, R. Kaibyshev, Microstructural evolution in a 5024 aluminum alloy processed by ECAP with and without back pressure, Mat. Sci. Eng. A 560 (2013) 178--192.
\bibitem{hump}F. Humphreys, M. Hatherly, Recrystallization and related annealing phenomena, 2004.
\bibitem{ma15}S. Malopheyev, S. Mironov, V. Kulitskiy, R. Kaibyshev, Friction-stir welding of ultra-fine grained sheets of Al--Mg--Sc--Zr alloy, Mat. Sci. Eng. A 624 (2015) 132--139.
\bibitem{m13}A. Mogucheva, E. Babich, B. Ovsyannikov, R. Kaibyshev, Microstructural evolution in a 5024 aluminum alloy processed by ECAP with and without back pressure, Mat. Sci. Eng. A 560 (2013) 178--192.
\bibitem{k05}R. Kaibyshev, K. Shipilova, F. Musin, Y. Motohashi, Continuous dynamic recrystallization in an Al--Li--Mg--Sc alloy during equal-channel angular extrusion, Mat. Sci. Eng. A 396 (2005) 341--351.
\bibitem{Kul}V. Kulitskiy, S. Malopheyev, Y. Buranova, S. Divinski, G. Wilde, R. Kaibyshev, Ultrafine-grained structure produced by FSW and ECAP in Al--Mg--Sc--Zr alloy, Mater. Sci. Forum 838 (2016) 379--384.
\bibitem{Har}L.G. Harrison, Influence of dislocations on diffusion kinetics in solids with particular reference to the alkali halides, Trans. Faraday Soc. 57 (1961) 1191--1199.
\bibitem{PD}A. Paul, T. Laurila, V. Vuorinen, S.V. Divinski, Thermodynamics, Diffusion and the Kirkendall Effect in Solids, Springer Int. Publ. Switzerland (2014).
\bibitem{Beke} D.L. Beke, I. G\"od\'eny, G. Erd\'elyi, F.J. Kedves, The pressure
dependence of grain-boundary diffusion of $^{65}$Zn in polycrystalline
aluminium, Phil. Mag. A 56 (1987) 673--680.

\bibitem{CINEB1} G. Henkelman, B. P. Uberuaga, and H J\'{o}nsson, A climbing image nudged elastic band method for finding saddle points and minimum energy paths, J. Chem. Phys. 113 (2000) 9901.
\bibitem{CINEB2} G. Henkelman and H. J\'{o}nsson, Improved tangent estimate in the nudged elastic band method for finding minimum energy paths and saddle points, J. Chem. Phys. 113 (2000) 9978-9985.
\bibitem{vtst} http://theory.cm.utexas.edu/vtsttools/
\bibitem{Kresse1993} G. Kresse and J. Hafner, Ab initio molecular dynamics for liquid metals, Phys. Rev. B 47 (1993) RC558.
\bibitem{Kresse1996} G. Kresse and J. Furthm\"{u}ller, Efficient iterative schemes for ab initio total-energy calculations using a plane-wave basis set, Phys. Rev. B 54 (1996) 11169.
\bibitem{PBE} J. P. Perdew, K. Burke, and M. Ernzerhof, Generalized Gradient Approximation Made Simple, Phys. Rev. Lett. 77 (1996) 3865.
\bibitem{Monkhorst1976} H. J. Monkhorst, and J. D. Pack, Special points for Brillouin-zone integrations, Phys. Rev. B 13 (1976) 5188.
\bibitem{Methfessel} M. Methfessel and A. T. Paxton, 
 High-precision sampling for Brillouin-zone integration in metals, Phys. Rev. B 40 (1989) 3616.
\bibitem{Cobulk}G.M. Hood, R.J. Schultz, J. Armstrong, Co tracer diffusion in Al, Phil. Mag. A 47 (1983) 775--779.
\bibitem{LC}A.D. Le Claire, Solute diffusion in dilute alloys, J Nucl. Mater. 69–70 (1978) 70--96.
\bibitem{D}S.V. Divinski, B.S. Bokstein, Recent advances and unresolved problems of grain boundary diffusion, Defect Diffusion Forum 309-310  (2011) 1--8.
\bibitem{Dasha}D. Prokoshkina, V. Esin, G. Wilde, S.V. Divinski, Grain boundary width, energy and self-diffusion in nickel: effect of material purity. Acta Mater. 61 (2013) 5188--5197.
\bibitem{SH}T. Surholt, Y. Mishin, Chr. Herzig, Grain-boundary diffusion and segregation of gold in copper - Investigation in the type-B and type-C kinetic regimes, Phys. Rev. B   50   (1994) 3577--3587.
\bibitem{G2}S.V. Divinski, G. Reglitz, G. Wilde, Grain boundary self-diffusion in polycrystalline nickel of different purity levels, Acta Mater. 58 (2010) 386--395.
\bibitem{Tokey}Z. Tokei, Z. Erdelyi, C. Girardeaux, A. Rolland, Effect of sulphur content and pre-annealing treatments on nickel grain-boundary diffusion in high-purity copper, Phil. Mag. A 80 (2000) 1075-1083.
\bibitem{Xavier}X. Sauvage, N. Enikeev, R. Valiev, Y. Nasedkina, M. Murashkin, Atomic-scale analysis of the segregation and precipitation mechanisms in a severely deformed AleMg alloy, Acta Mater. 72 (2014) 125--136.
\bibitem{Dillon}S.J. Dillon, M. Tang, W.C. Carter, M.P. Harmer, Complexion: A new concept for kinetic engineering in materials science, Acta Mater. 55 (2007) 6208--6218.
\bibitem{N2}A.A. Nazarov, Kinetics of grain boundary recovery in deformed polycrystals, Interface Sci. 8 (2000) 315-322.
\bibitem{Hoesler}A. H\"a{\ss}ner, Untersuchung der Korngrenzendiffusion von Zn-65 in $\alpha$-Aluminium--Zink--Legierungen, Kristall und Technik 9 (1974) 1371-1388.
\bibitem{Schafler}M.B. Kerber, M.J. Zehetbauer, E. Schafler, F.C. Spieckermann, S. Bernstorff, T. Ungar, X-ray line profile analysis - An ideal tool to quantify structural parameters of nanomaterials, JOM 63 (2011) 61--69.
\bibitem{Geise}S.V. Divinski, J. Geise, E. Rabkin, Chr. Herzig, Grain Boundary Self-Diffusion in $\alpha$-Fe of Different Purity: Effect of Dislocation Enhanced Diffusion, Z. Metallkunde 95 (2004) 945--952.
\bibitem{ZB}R. Swiatek, M. Zehetbauer, B. Mikulowski, Work hardening by deformation induced vacancies in low temperature deformed aluminium single crystals, Mat. Sci. Eng. A 234 (1997) 441--444.
\bibitem{Sommer}J. Sommer, Chr. Herzig, Direct determination of grain-boundary and dislocation self-diffusion coefficients in silver from experiments in type-C kinetics, J. Appl. Phys. 72 (1992) 2758--2766.
\bibitem{DHK1}S. V. Divinski, F. Hisker, Y.-S. Kang, J.-S. Lee, Chr. Herzig, $^{59}$Fe Grain Boundary Diffusion in Nanostructured $\gamma$-Fe-Ni. Part I: Radiotracer Experiments and Monte-Carlo Simulation in Type-A and B Kinetic Regimes, Z. Metallkde. 93 (2002) 256--265.
\bibitem{DHK2}S. V. Divinski, F. Hisker, Y.-S. Kang, J.-S. Lee, Chr. Herzig, $^{59}$Fe Grain Boundary Diffusion in Nanostructured $\gamma$-Fe-Ni. Part II: Effect of Bimodal Microstructure on Diffusion Behaviour in Type-C Kinetic Regime, Z. Metallkde. 93 (2002) 265--279.
\bibitem{RK}L. Klinger, E. Rabkin, Beyond the Fisher model of grain boundary diffusion: Effect of structural inhomogeneity in the bulk, Acta Mater. 47 (1999) 725--734.
\bibitem{Szabo}I.A. Szabo, D.L. Beke, F.J. Kedves, On the transition between the C and B Kinetic regimes for grain-boundary diffusion, Phil. Mag. 62 (1990) 227--239.
\bibitem{marq}E.A. Marquis, D.N. Seidman, Nanoscale structural evolution of Al$_3$Sc precipitates in Al(Sc) alloys, Acta Mater. 49 (2001) 1909--1919.
\bibitem{murr}J.L. Murray, The Al--Mg (aluminum--magnesium) system, Bulletin of Alloy Phase Diagrams 3 (1982) 60--74.
\bibitem{harada}Y. Harada, D.C. Dunand, Microstructure of Al$_3$Sc with ternary transition-metal additions, Mat. Sci. Eng. A 329 (2002) 686--695.
\bibitem{Eshelby}J.D. Eshelby, The determination of the elastic field of an ellipsoidal inclusion, and related problems, Proc. Royal Soc. London A 241 (1957) 376--396. 
\bibitem{Ankit}A. Gupta, B. Tas Kavakbasi, B. Dutta, B. Grabowski, M.Peterlechner, T. Hickel, S.V. Divinski, G. Wilde, J. Neugebauer, Low temperature features in the heat capacity of unary metals and intermetallics for the example of bulk aluminum and Al$_3$Sc, Phys Rev B (2017) submitted.
\bibitem{LCL}A.D. Le Claire, A.B. Lidiard, Correlation effects in diffusion in crystals, Phil. Mag. 47 (1956) 518--527.
\bibitem{Mog3}A. Mogucheva, D. Yuzbekova, T. Lebedkina, M. Lebyodkin, R. Kaibyshev, Influence of Severe Plastic Deformation on Mechanical Properties of an AA5024 Alloy, Mater. Science Forum 879 (2017) 1317--1322.
\bibitem{Kalnins}J.R. Kalnin, E.A. Kotomin, J. Maier, Calculations of the effective diffusion coefficient for inhomogeneous media, J. Phys. Chem. Solids 63 (2002) 449--456.
\bibitem{BM}I.V. Belova, G.E. Murch, Diffusion in nanocrystalline materials, J. Phys. Chem. Solids,
64 (2003) 873--878.

\end{thebibliography}
\end{document}